\RequirePackage{fix-cm}
\documentclass[smallextended]{svjour3}       
\smartqed  
\usepackage{graphicx}
 \usepackage{caption}
 \usepackage{float}
 \usepackage[linesnumbered, ruled]{algorithm2e}
\begin{document}

\title{Packing Unequal Circles into a Square Container by Using Partitioning Narrow Action Spaces and Circle Items}

\titlerunning{Global Optimization}        

\author{Kun He\and Mohammed Dosh*\and Shenghao Zou}

           \institute{K. He  \at {Department of Computer Science and Technology,  Huazhong University of Science \& Technology, Wuhan 430074, China.}
                   \\\email{brooklet60@hust.edu.cn}
           \and
                   M. Dosh \footnote \\ \at {Department of Computer Science and Technology, Huazhong University of Science \& Technology, Wuhan 430074, China. \and Faculty of education for girls, Kufa University, Najaf, Iraq.}
                \\ \email {i201522007@hust.edu.cn, mohammedh.dosh@uokufa.edu.iq}
           \and
                  S. Zou   \at {Department of Computer Science and Technology, Huazhong University of Science \& Technology, Wuhan 430074, China.}
              \\ \email{shenghaozou@hust.edu.cn}
                  }

\date{Received: date / Accepted: date}

\maketitle

\begin{abstract}
We address the NP-hard problem of finding a non-overlapping dense packing pattern for \textit{n} Unequal Circle items in a two-dimensional Square Container (PUC-SC) such that the size of the container is minimized. Based on our previous work on Action Space-based Global Optimization (ASGO) that approximates each circle item as a square item to find large unoccupied spaces efficiently, we propose an optimization algorithm based on the Partitioned Action Space and Partitioned Circle Items (PAS-PCI). The PAS is used to partition the narrow action space on the long side to find two equal action spaces to fully utilize the unoccupied spaces. The PCI are used to partition the circle items into four groups based on item sizes for the basin-hopping strategy. Experiments on two sets of benchmark instances show the effectiveness of the proposed method. In comparison with our previous ASGO algorithm on the 68 tested instances published with ASGO, PAS-PCI not only achieves smaller containers in 64 instances and matches the other 4 but also runs faster in most instances. In comparison with the best record of the Packomania website for 94 instances, PAS-PCI finds smaller containers for 82 and matches the other 12. Note that we updated 19 records (47-48, 51-54, 57, 61-72) that had remained unchanged since 2013.
\end{abstract}

\section{Introduction}
\label{sec1}
The aim of single container packing problem is to find the densest possible packing patterns for the given items in a single container. The container can be square, circular, semi-circular, polygonal, cube-shaped or rectangular, and the items can be rectangles, circles or irregularly shaped. As an important class of optimization problems, packing problems have numerous applications in industry and academia, such as applied mathematics, material manufacturing, material cutting, logistics, wireless communication, the fashion industry, architecture layout, and the motor cycle industry. As an NP-hard problem, however, there is no exact algorithm to obtain optimality in polynomial time unless N = NP, and researchers have resorted to heuristics or approximation methods.

We consider a two-dimensional Circle Packing Problem (CPP) in which items are circles. The CPP has been well studied in a wide range of papers. Hifi and M'Hallah \cite{9} reviewed the most relevant literature on efficient models and methods for packing circular items in different types of containers. The CPP is classified into two categories based on whether the circle items are equal \cite{21,29,32} or unequal  \cite{1,2,12}. Other variants exist that consider additional constraints, such as the CPP with equilibrium constraints \cite{8,18}.

For the equal-circle packing problem, the most common approaches are based on the quasi-physical method. Lubachevsky and Graham \cite{24} evaluated each item as a rigid billiard, viewed its movement under a collision force, and proposed a billiards simulation algorithm based on the collision forces between the objects.
They also proved that their algorithm could obtain an optimal solution if \textit{n}, the number of circles, equals $3{k}({k}+1)+1$ for any positive integer \textit{k}. Huang and Ye \cite{13,14} proposed a quasi-physical global optimization method that considered each circle item an elastic object and considered two types of movement, a smooth movement driven by elastic forces and a violent movement driven by strong repulsive and attractive forces. Liu et al. \cite{19} proposed another quasi-physical algorithm incorporating a new pattern update mechanism with an improved Energy-Landscape-Paving method. Other researchers applied meta-heuristics to solve the CPP. For example, Grosso et al. \cite{4} proposed a genetic algorithm for packing equal circles in a circular container based on the Monotonic Basin-Hopping approach and a population-based variant basin-hopping strategy.

For the unequal-circle packing problem, there are two main categories: construction methods and global optimization methods. The construction methods start by constructing a pattern by placing circles one by one according to specific rules. Some researchers fix the container size and rearrange the positions of the circles through each constructive procedure, whereas others adjust the container radius and container centre through the constructive procedure. Huang et al. \cite{15} proposed two greedy algorithms, B1.0 and B1.5. B1.0 placed the circles one by one according to the maximum-hole degree of the current placement, and B1.5 improved B1.0 with a self-look-ahead search strategy. L\"{u} and Huang \cite{25} incorporated a Pruned Enriched Rosenbluth Method (PERM) using the maximum-hole degree strategy, in which the PERM strategy was to efficiently prune and enrich branches and the maximum-hole degree was defined to evaluate the benefit of a partial pattern. Hifi and M'Hallah \cite{10} proposed a three-phase algorithm: a dynamic search phase that successively packed the ordered set of circles, an adaptive phase that used intensification and diversification to find a smaller container, and a restarting phase based on the hill climbing approach. Kubach, Bortfeldt and Gehring \cite{17} formulated several greedy algorithms and parallelized these algorithms by using a master-slave approach followed by a subtree distribution model.

Global optimization methods for solving the unequal-circle packing problem concentrate on improving the pattern iteratively rather than constructing a good initial pattern. Hi et al. \cite{11} presented a Simulated Annealing (SA) approach that defined an energy function for local optimization by using certain pattern transformation methods. Stoyan and Yaskov \cite{28} utilized a combination of tree search and reduced gradient methods for local optimization and translated from one local minimum to another local minimum based on the concept of active inequalities and the Newton method. Fu et al. proposed an Iterated Tabu Search approach (ITS) \cite{3} starting from randomly generated solutions, and a perturbation operator was subsequently employed to reconstruct the incumbent solution. Zeng et al. \cite{31} presented a Tabu Search and Variable Neighbourhood Descent (TS-VND), which is an adaptation of the tabu search procedure of ITS algorithms \cite{3}. Lopez and Beasley \cite{22} viewed this problem as scaling the radii of the unequal circles so that all of them could be packed into a fixed-size container; they improved the pattern quality by using a perturbation strategy of swapping two circles. Additionally, in 2016, they used a new and merged metaheuristic for packing unequal circles in a fixed-size container \cite{23}. Specht \cite{26} proposed a novel algorithm to analyse the unoccupied areas in a packing pattern by graph theoretical methods. It used the object-jumping strategy to achieve densification. This strategy is efficient when the smallest jammer is smaller than the largest unoccupied space.

In this paper, we address an important version of the Circle Packing Problem that attempts to find a non-overlapping dense Packing pattern for \textit{n} Unequal Circle items in a two-dimensional Square Container (PUC-SC) such that the size of the container is minimized. We propose a Partitioning Narrow Action Space and Circle Items (PAS-PCI) algorithm for solving the PUC-SC. The proposed method is inspired by our previous work. In 2011 and 2012, He et al. defined the concept of an action space for the rectangular packing problem \cite{6,7}, and each maximal unoccupied rectangular space is called an action space in a packing pattern. In 2015, He et al.\cite{5} proposed an Action Space based Global Optimization (ASGO) algorithm for the problem that packs unequal circles into a square container. For some given patterns, ASGO utilized the Limited-memory Broyden-Fletcher-Goldfarb-Shanno (LBFGS) algorithm \cite{20} for continuous optimization to reach the local minimums. Then, each circular item is approximated as a square item, and an action-space-based basin-hopping strategy is adapted to find the larger unoccupied spaces as candidate places to replace some of the most overlapping items to jump from a local minimum.

Specifically, PAS-PCI uses a new basin-hopping strategy based on our observation that the small circles and the large unoccupied circle are robustly complementary to each other. We initially partition the circles into several groups according to their sizes. When reaching a packing pattern with the local minimal potential energy, we select items with the maximal deformation from each group and move them to the best-matching or randomly chosen unoccupied action spaces. For the narrow action spaces, we place two items selected from one or two adjacent groups, indicating that they are similar sizes, in the centre of the two partitioned action spaces, which we call neighbour spaces. The new basin-hopping strategy helps reduce local cycling and pushes the search to a promising area.

PAS-PCI also applies a perturbation operator to escape from the local minimum, which is obtained using LBFGS continuous optimization and the basin-hopping strategy. The perturbation operator swaps two similar-size circles or two randomly selected circles. Finally, a post-processing procedure is applied to reduce the container size further when a feasible packing pattern is reached. We perform computational experiments on two sets of benchmark instances and demonstrate the effectiveness of the proposed method by comparing it with ASGO \cite{5}, the current best algorithm published in the literature, and the best-known results downloaded from the Packomania website \cite{27}.

\section{Problem formulation and the general quasi-physical approach}
\label{sec2}
For the PUC-SC problem, we are given $n ({n} \in {N^+})$ circles $C_1, C_2,\dots , C_n$, with integer radii $r_1, r_2, r_3, \dots, r_n$, and asked to find the smallest square container of size \textit{L} such that all circle items are packed feasibly, i.e., without any overlapping between any pair-wise circles $(C_i \bigcap C_j =\phi )$ and with all circles fitting completely inside the container.

\begin{figure}[htbp]
 \centering
 \includegraphics[scale=0.4]{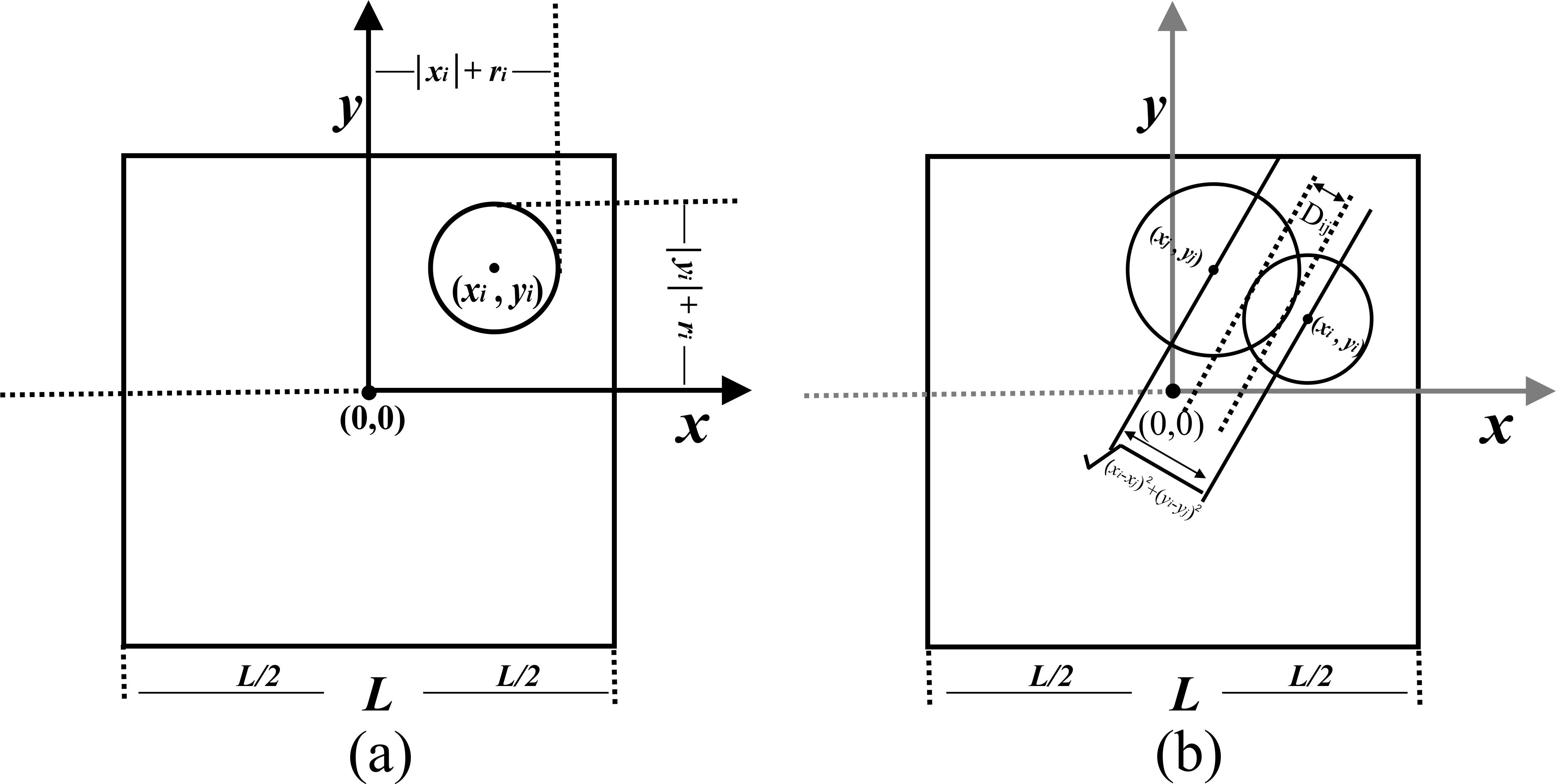}
  \caption {Denotations for the PUC-SC problem.}
 \label{fig:key1}
 \end{figure}

Specifically, let the centre of the square container be located at the origin of a two-dimensional Cartesian coordinate system. Denote the centre coordinate of circle $C_i (1 \leq i \leq n)$ as $(x_i, y_i)$, as shown in Fig. \ref{fig:key1}a. We define a packing pattern as $X = (x_1, y_1, \dots, x_i, y_i, \dots, x_n, y_n)$. Let $ \sqrt{(x_i-x_j )^2+(y_i-y_j )^2}$ indicate the Euclidean distance between the centres of circle $C_i$ and $C_j$ (Fig. \ref{fig:key1}b). The PUC-SC problem can be formulated as follows.

        min  \textit{L}\\
s.t.    (1)  $ max(|x_i |,|y_i | )\leq(L/2-r_i),    1 \leq i \leq n.$

\indent~(2)  $ \sqrt{((x_i-x_j )^2+( y_i-y_j)^2 )}  \geq (r_i+r_j),     1 \leq i < j \leq n.$
        
Here, (1) imposes the condition that any circle should not be extended outside the container, and (2) imposes no overlap between any pair-wise circles.

The quality of a feasible solution is measured by the size of the square container \textit{L}. We start from the current known best container size $L_{best-current-result}$ published by ASGO, and in the case when a feasible solution is found, we will try to reduce the container size and seek a better solution.

For a fixed container size \textit{L}, we adopt the quasi-physical approach \cite{30}, considering all \textit{n} circles smooth elastic solids and the square container \textit{L} a hollow rigid solid. Initially, the circles are randomly placed; then, they will move according to the forces generated by the deformation. For a pattern (\textit{X}, \textit{L}), the overlapping between any pair-wise solids will cause elastic deformation and generate potential energy. The overlapping depth between circle $C_i$ and the vertical border is as follows:
  \begin{equation}
  \label{eq1}
  D_{iv}=max(r_i+|x_i |- (L/2),0 ).
  \end{equation}
However, the overlapping depth between circle $C_i$ and the horizontal border is as follows:
  \begin{equation}
  \label{eq2}
  D_{ih}=max(r_i+|y_i |-(L/2),0 ).
  \end{equation}
The embedding depth of two circles $C_i$ and $C_j$ is as follows:
   \begin{equation}
   \label{eq3}
   D_{ij}= max ((r_i  + r_j)- \sqrt{(x_i-x_j )^2+(y_i-y_j )^2},0 )
   \end{equation}
Then, we compute the total elastic potential energy for pattern (\textit{X}, \textit{L}) as the penalty function \cite{5}:
   \begin{equation}
   \label{eq4}
   U_e (X) = \sum_{i=1}^{n-1}\sum_{j=i+1}^{n}(D_{ij}^2)+\sum_{i=1}^{n}(D_{iv}^2 + D_{ih}^2).
   \end{equation}

$U_e$(\textit{X}) measures the extent of overlap and is clearly nonnegative. If $U_e$(\textit{X}) = 0, then we reach a feasible solution. If $U_e({X}) > 0$, then overlap exists.

When the circles movement reaches a local minimum of the penalty function, the circles can no longer move even when some overlap exists. We define the pain degree \cite{5}for a circle \textit{i} in Eq. \ref{eq5}, where $D_{iv}$ and $D_{ih}$ indicate the overlapping depth to the container borders vertically and horizontally, and $D_{ij}$ indicates the overlapping depth to circle item \textit{j}.
   \begin{equation}
   \label{eq5}
P_i  =(D_{iv}^2+D_{ih}^2+\sum_{j=1 j\neq i}^{n}D{ij}^2)/r_i^2
   \end{equation}

We then pick a circle $C_i$ with the largest pain degree, i.e., the most squeezed circle, and move it to an unoccupied place. This approach is the basin-hopping strategy that drives the search to a promising area. Then, the physical movement can continue; we run the above process iteratively until we find a feasible solution or the iterative process reaches a predefined limit.

\section{PAS-PCI approach}
\label{sec3}
\begin{figure}[htbp]
\centering
\includegraphics[scale=0.4]{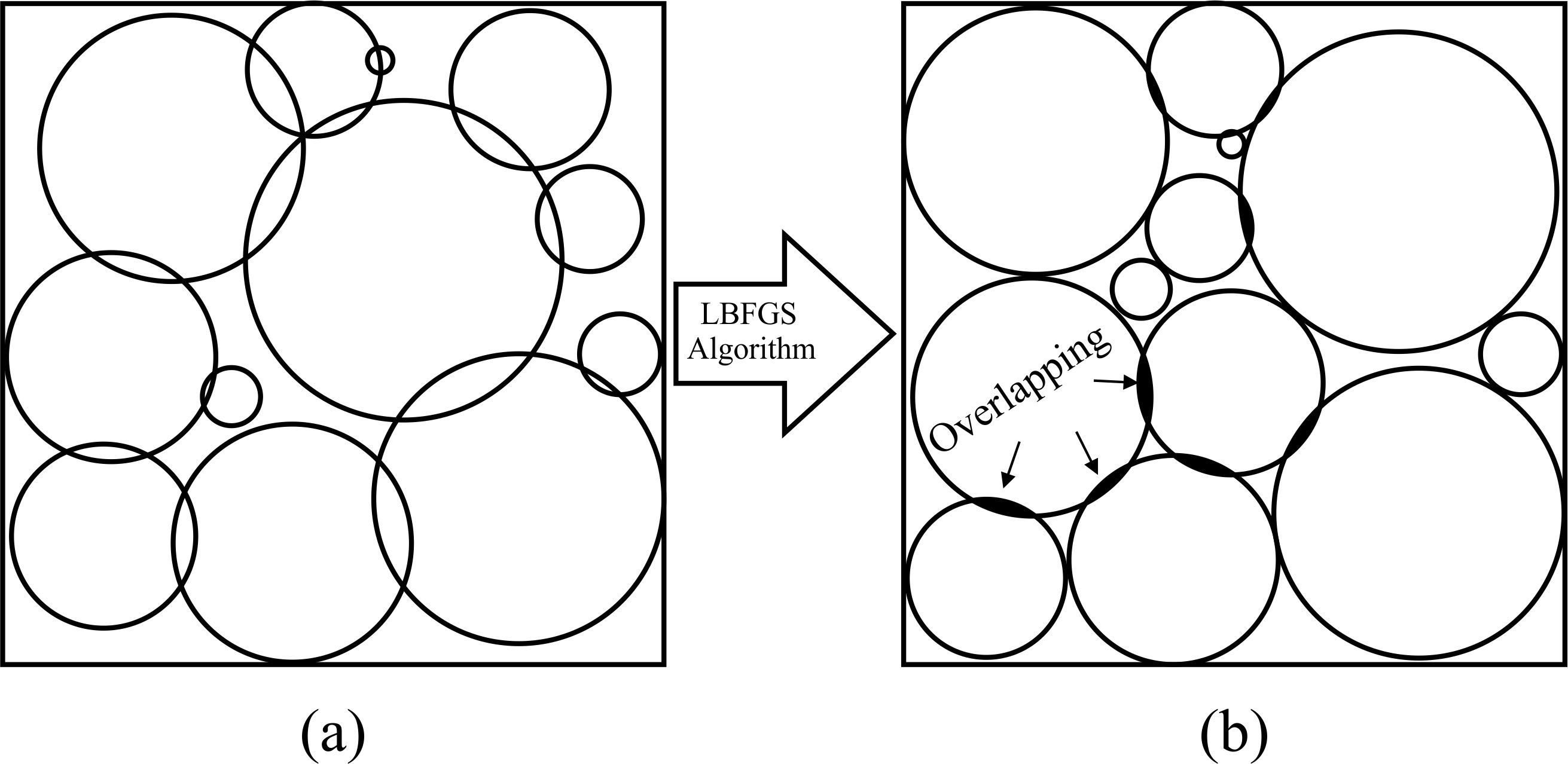}
\caption {Fig. 2. Applying LBFGS to a pattern.}
\label{fig:key2}
\end{figure}

The Limited-memory BFGS (LBFGS) is in the family of the Quasi-Newton method and is usually used to solve large nonlinear optimization problems. For a fixed container size $L_0$, initialized by the current best record for each instance, we randomly place the circle items inside the container to generate multiple patterns. For each pattern, we apply the LBFGS algorithm for continuous optimization until a local minimum is reached. For example, Fig. \ref{fig:key2}a shows a random pattern, and Fig. \ref{fig:key2}b shows the new pattern in the local minimum after applying LBFGS, in which overlaps remain. We then use a basin-hopping strategy based on the Partitioned Action Space and Partitioned Circle to jump out of the stuck pattern.

\subsection{Basin-hopping strategy}
\label{sec3:1}
Researchers have proposed many basin-hopping methods to help the stagnated pattern escape from the local optimum by learning the advantages from the search history. An intuitive strategy is to transfer the most squeezed circle to a random place inside the container; however, this method is not effective when the search is located in a large space. Another popular strategy is to swap two circles by various rules, such as swapping pair-wise circles of similar sizes. Although the second strategy is not efficacious for packing equal circles, it is very useful for packing unequal circles. In this paper, we propose a new basin-hopping strategy by partitioning the narrow action spaces and partitioning circle items according to their sizes.

\begin{definition}[Partitioned Circle Items, PCI]
\label{def1}
For a pattern (\textit{X}, \textit{L}) with \textit{n} circle items, we partition the items into four sets, $S_1$ to $S_4$, based on their radii. Without loss of generality and assuming the circles are numbered from 1 to \textit{n} from small to large, we then obtain [1, int (\textit{n}/4)], [int (\textit{n}/4) +1, int (\textit{n}/2)], [int (\textit{n}/2) +1, int (3\textit{n})/4], and [int (3\textit{n}/4) +1, \textit{n}].
\end{definition}

\begin{figure}[htbp]
\centering
\includegraphics[scale=0.45]{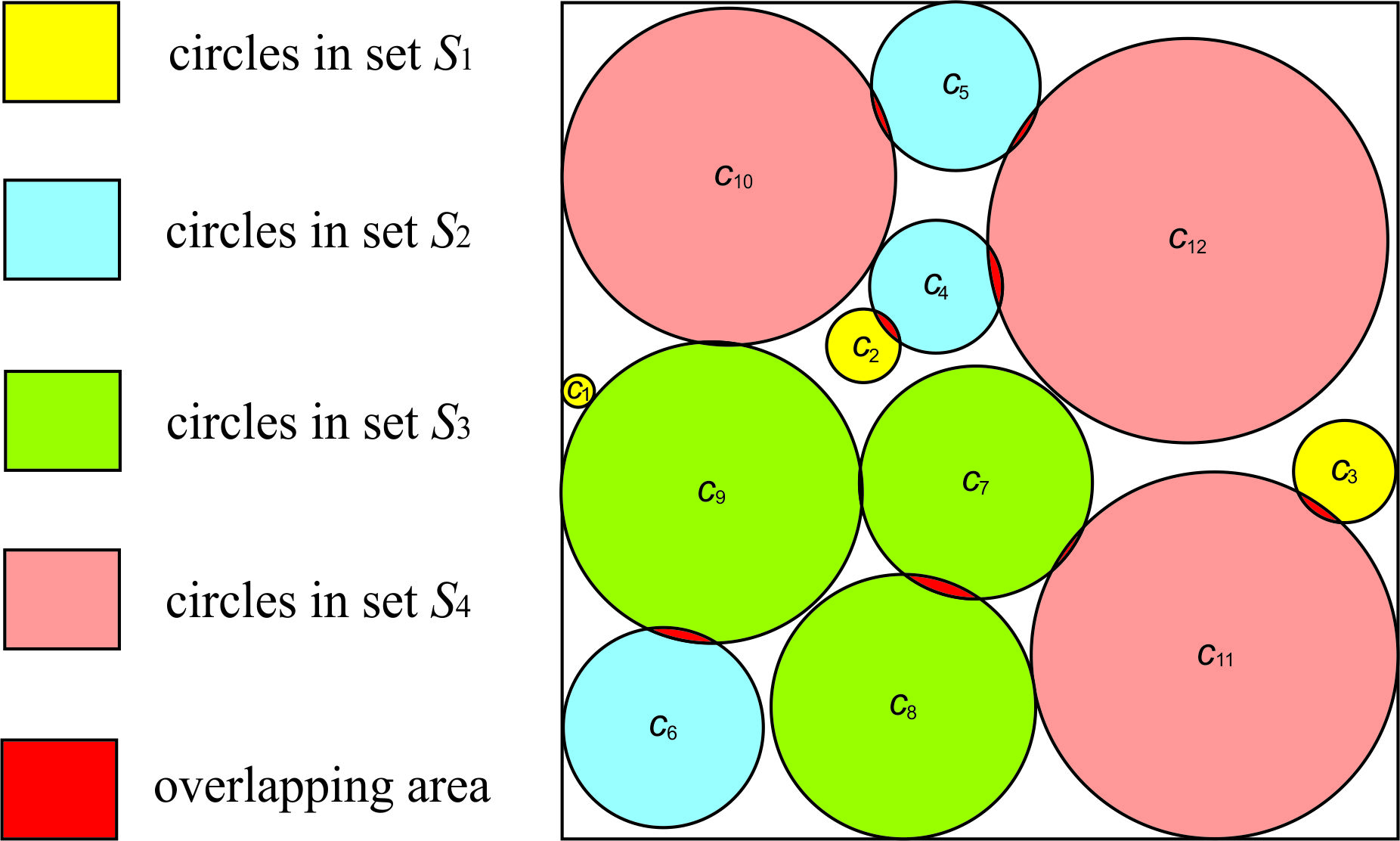}
\caption {An example of the partitioned sets for \textit{n} = 12. }
\label{fig:key3}
\end{figure}

Fig. \ref{fig:key3} shows an example for \textit{n} = 12. $S_1$ contains the smallest items, and $S_4$ contains the largest.

The locations of large circles in the container are fixed to some extent, whereas the small circles are relatively flexible. Therefore, we give circles in $S_1$ and $S_2$ more priority to move to the Best Matching Spaces (BMS) or other suitable unoccupied spaces and drive the search to a promising area.

\begin{definition}[Best Matching Space, BMS]
\label{def2}
The best matching action space of an item is an action space in which the length of the short side is closest to the diameter of the circle.
\end{definition}

For a stuck pattern (\textit{X}, \textit{L}), we must discover and measure the unoccupied spaces to move a squeezed circle to the right place, but it is difficult to perform accurate calculations because the unoccupied spaces are irregularly shaped. Some researchers pseudo-place a number of additional check disks randomly in the container \cite{16} and calculate the elastic energy of the check disks to evaluate the size of the unoccupied spaces, which is effective for packing equal circles. However, this strategy might not find all of the useful occupied space due to the random placement of the check disks, and some check disks may be placed inside some circle items.

The key issue for the basin-hopping strategy is how to find all of the large unoccupied spaces effectively. We borrow the action space concept from the rectangular packing problem to find the unoccupied spaces. For the rectangular packing problem, an action space is an unoccupied rectangular space into which a dummy rectangle could be feasibly placed, and each edge of the dummy rectangle touches at least one item or the container \cite{6,7}. By approximating the circles to square items, He et al. proposed a novel approach to find all unoccupied spaces as action spaces \cite{5}. The action spaces could approximately measure the sizes of the corresponding unoccupied spaces, and the action spaces could be calculated quickly. We adopt the action space method to find unoccupied spaces, and we consider additional characteristics of the action spaces to fully utilize the unoccupied spaces.

\begin{figure}[htbp]
\centering
\includegraphics[scale=0.45]{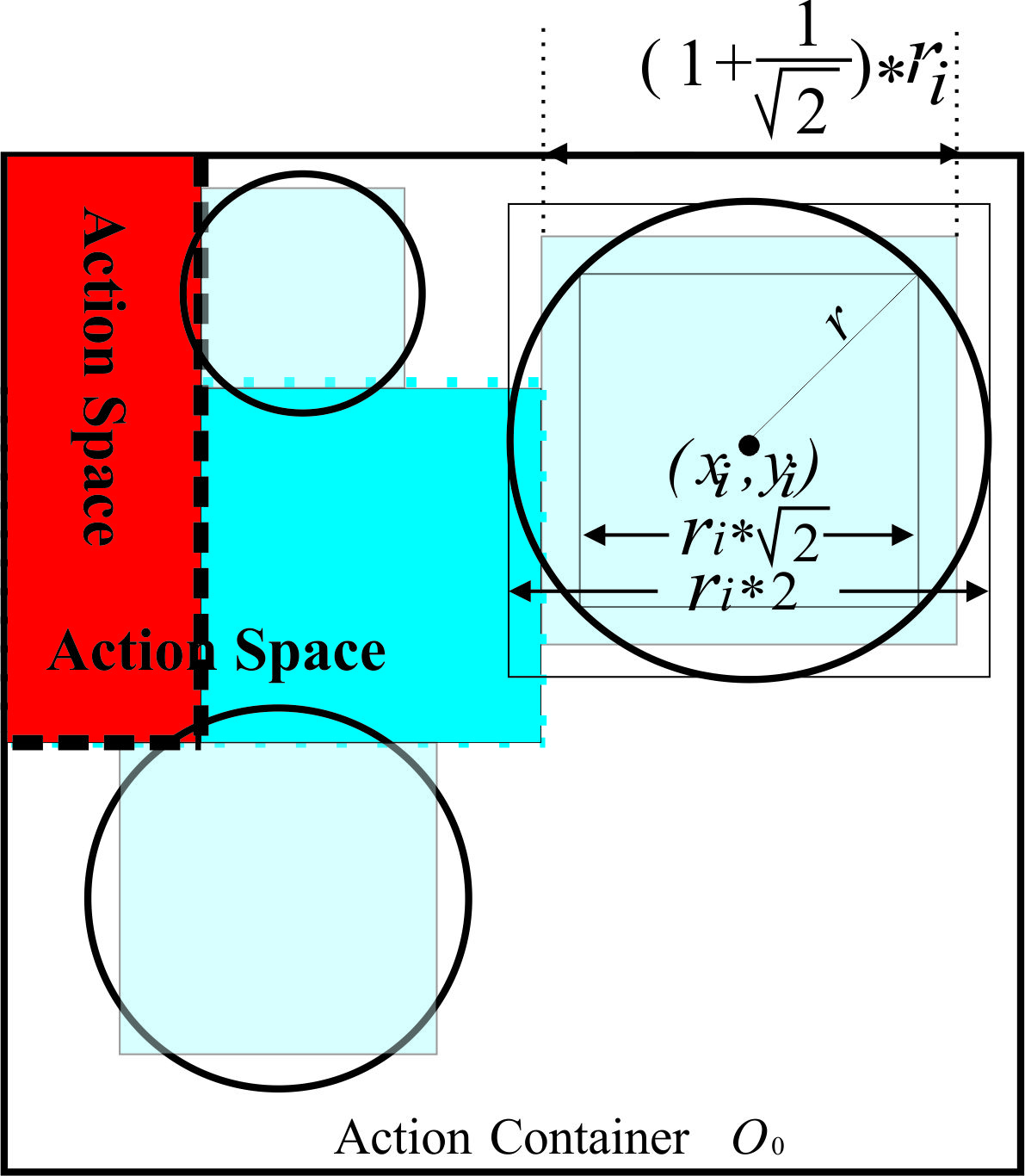}
\caption {Approximate each circular item as a square item.}
\label{fig:key4}
\end{figure}

\begin{definition}[Square container $O_0$ and square item $O_i$]
\label{def3}
The square container $O_0$ is only the container for the circle-packing problem. We then approximate each circular item $C_i (i \in [1, n])$ as a square item $O_i$ by setting its width as $(1+(1/\sqrt{2})) r_i$, as shown in Fig. \ref{fig:key4}.
\end{definition}

For a stuck pattern, we approximate each circular item as a square item and adapt the idea of the action space to find all large unoccupied spaces. In the beginning, there is only one action space, the container $O_0$. After placing all of the square items into the action spaces, the action space list is updated. If a square item coincides completely with an action space, then no more action space can be created. If a square item intersects and coincides with an action space in three dimensions, then one new action space is created. If the item coincides with an action space in two dimensions, then two new action spaces are created.

For an action space \textit{i}, denote its width and height as $w_i$ and $h_i$, and denote the coordinates of its left-bottom vertex and right-top vertex as $(x_{i1}, y_{i1})$ and $(x_{i2}, y_{i2})$, respectively. We measure the size of an action space in two ways, the length of the short side and the perimeter, and sort all action spaces in non-ascending order based on the two metrics to create two lists. Let $l_1$ and $l_2$ are two lists, and each list contains the top 10 action spaces, respectively.
 (i) Lexicographic order (list $l_1$): $\min(h_i, w_i) \geq \min(h_j, w_j)$ for \textit{i} in front of \textit{j}.
 (ii) Traditional order (list $l_2$): $h_i + w_i \geq h_j + w_j$ for \textit{i} in front of \textit{j}.

\begin{figure}[htbp]
\centering
\includegraphics[scale=0.4]{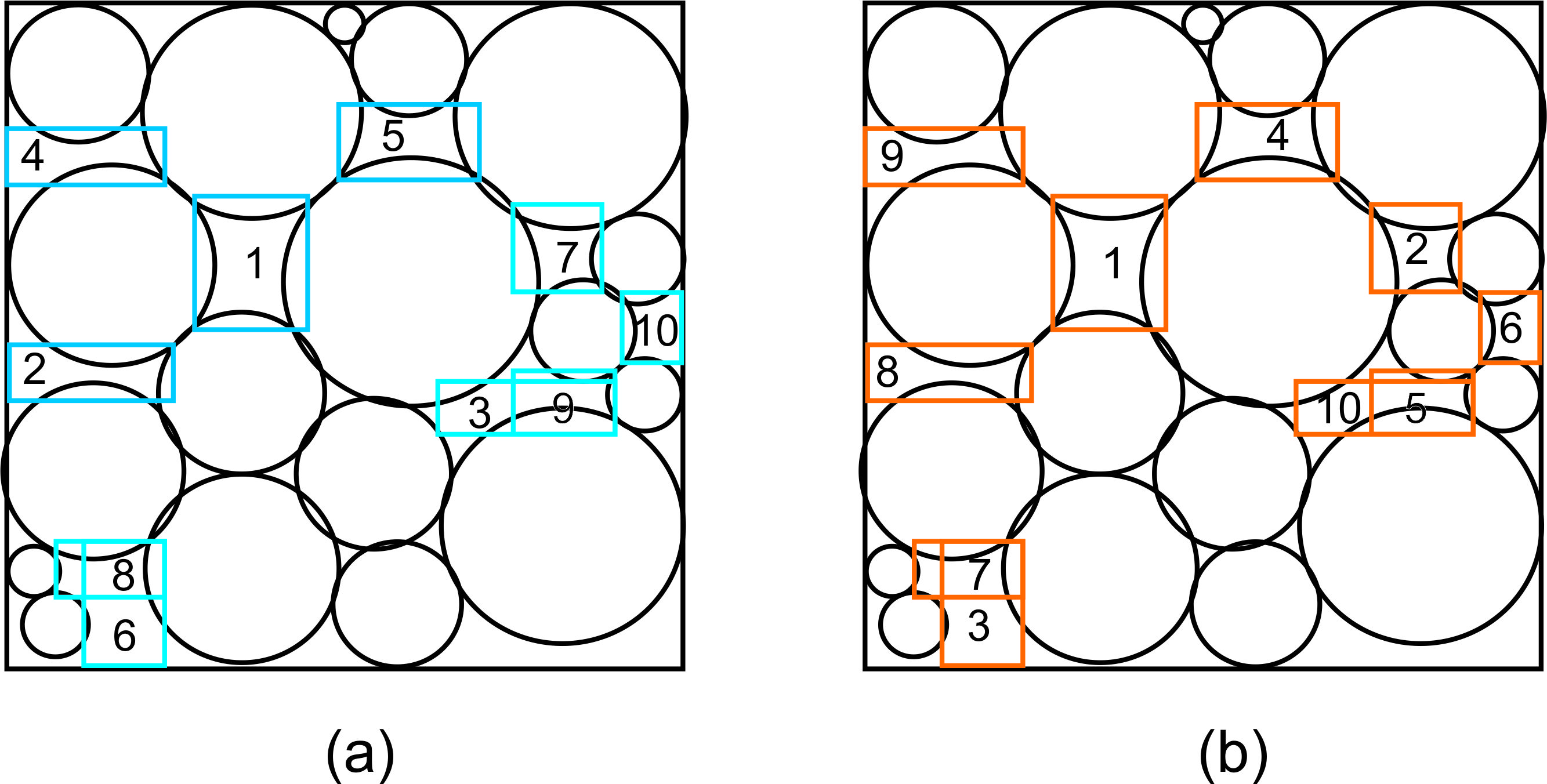}
\caption { List of action spaces obtained by two methods.}
\label{fig:key5}
\end{figure}

There are more narrow action spaces in list $l_2$, and the narrow action spaces are sorted towards the front compared with their positions in list $l_1$. Thus, we can find more suitable action spaces for moving squeezed circles, particularly for small circles. Fig. \ref{fig:key5}a shows the action spaces in list $l_1$, and Fig. \ref{fig:key5}b shows the action spaces in list $l_2$.

\begin{definition}[Narrow Action Space, NAAS]
\label{def4}
An action space \textit{i} is called narrow if the long side is at least double the short side, i.e., either $h_i \leq 0.5 w_i$, or $w_i \leq 0.5 h_i$.
\end{definition}

\begin{figure}[htbp]
\centering
\includegraphics[scale=0.45]{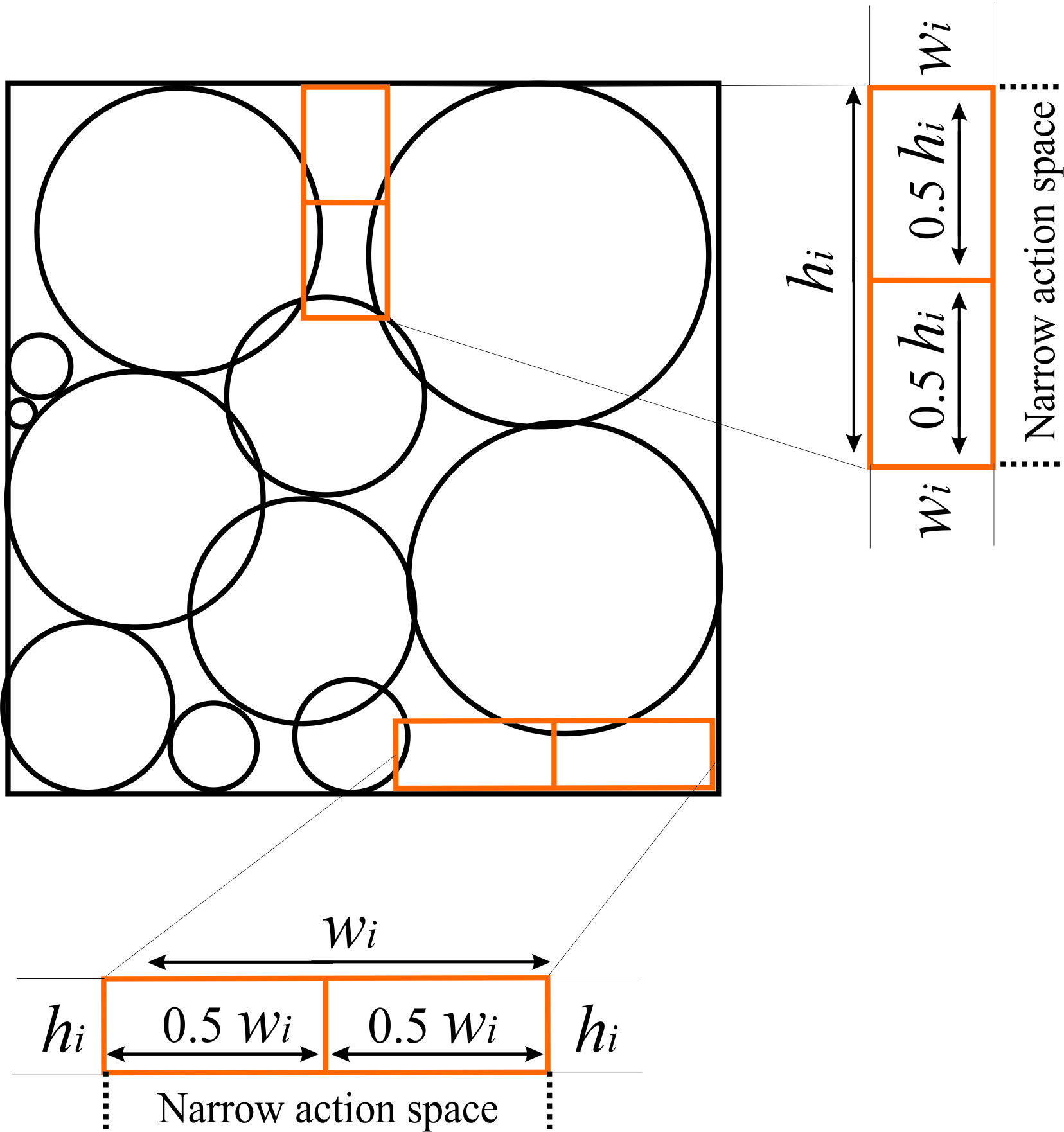}
\caption {Partition the narrow action spaces. }
\label{fig:key6}
\end{figure}

\begin{definition}[Neighbour Action Spaces, NEAS]
\label{def5}
We partition all narrow action spaces in lists $l_1$ and $l_2$ by dividing each Narrow Action Space at the centre of the long side to obtain two equal-size action spaces, which we call Neighbour Action Spaces, as shown in Fig. \ref{fig:key6}.\\
\end{definition}

\begin{definition}[Neighbour Space Occupying, NSO]
\label{def6}
Neighbour Space Occupying (NSO) is the placement of two items in the centres of the Neighbour Action Spaces.
\end{definition}

\begin{figure}[htbp]
\centering
\includegraphics[scale=0.4]{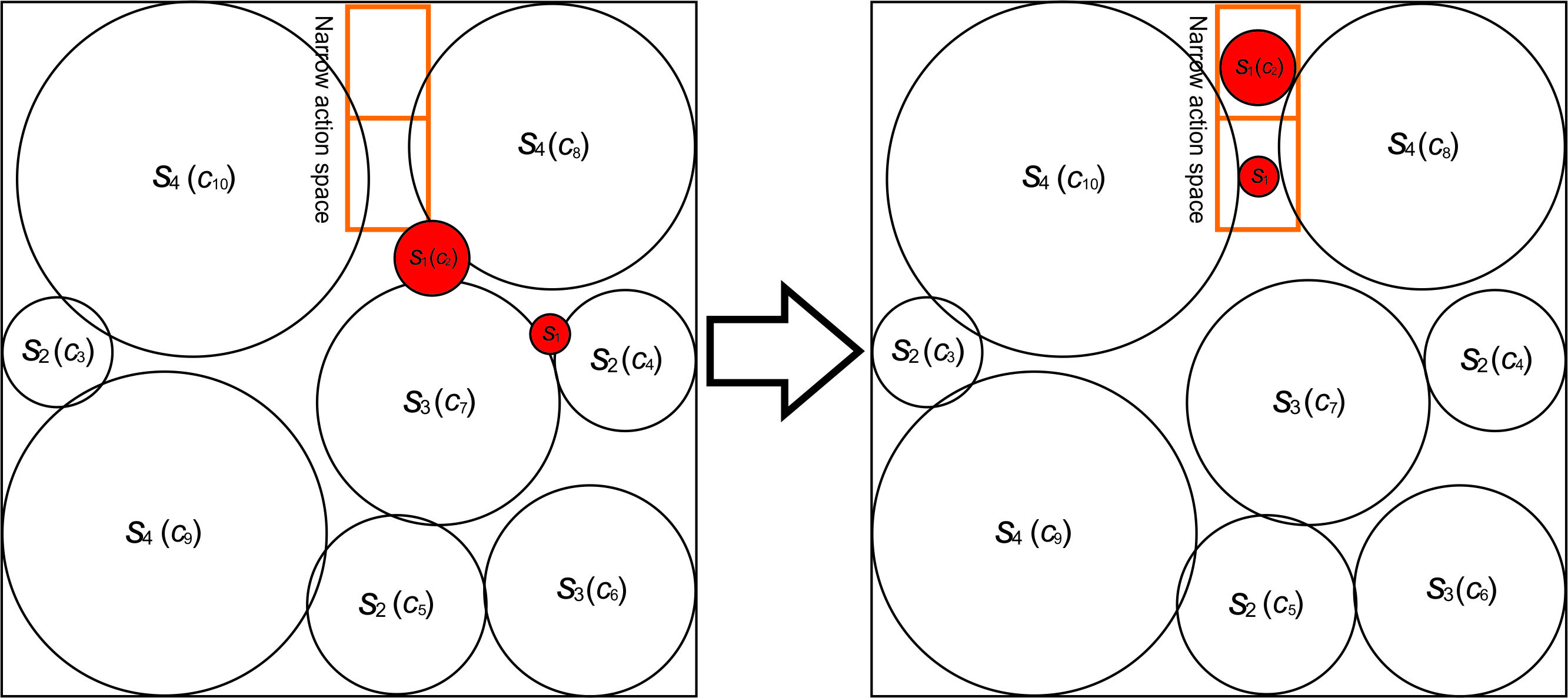}
\caption { Move two small items to the centres of the Neighbour Action Space.}
\label{fig:key7}
\end{figure}

For a narrow action space, we select two items with the greatest pain degree from $S_1$, or one from $S_1$ and the other from $S_2$, and perform NSO. Fig. \ref{fig:key7} shows an example.

\subsection{Algorithm description}
\label{sec3:2}
The PAS-PCI algorithm includes three phases: the initial continuous optimization phase, iterative processing phase and post-processing phase. In the initial continuous optimization phase, we randomly generate a number of packing patterns and use LBFGS for continuous optimization to reach patterns with the local minimum potential energy. Then, in the iterative processing phase, we use basin hopping and perturbation operators together with LBFGS to find a feasible pattern. We initially run the basin hopping and LBFGS iteratively for at most 20 iterations. The basin-hopping algorithm uses intelligent strategies to jump out of the local minimum trap, whereas LBFGS is applied to reach new local minimum patterns. Then, the perturbation operator strategy is used to perturb the patterns to some extent before we run 20 iterations of the basin hopping plus LBFGS once again. The outer iterations for perturbation will run for 5 loops. The entire iterative processing phase will stop once a feasible pattern is found. Finally, we use the post-processing algorithm to reduce further the container size for a feasible pattern. For the parameters, we refer to the ASGO algorithm and make some tiny changes. Fig. \ref{fig:key8} illustrates the framework of the PAS-PCI algorithm.

In the following subsections, we will initially discuss the entire framework and then the details for the basin hopping and post-processing strategies.

\subsubsection{PAS-PCI framework}
\label{sec3:2:1}

\begin{figure}[htbp]
\centering
\includegraphics[scale=0.5]{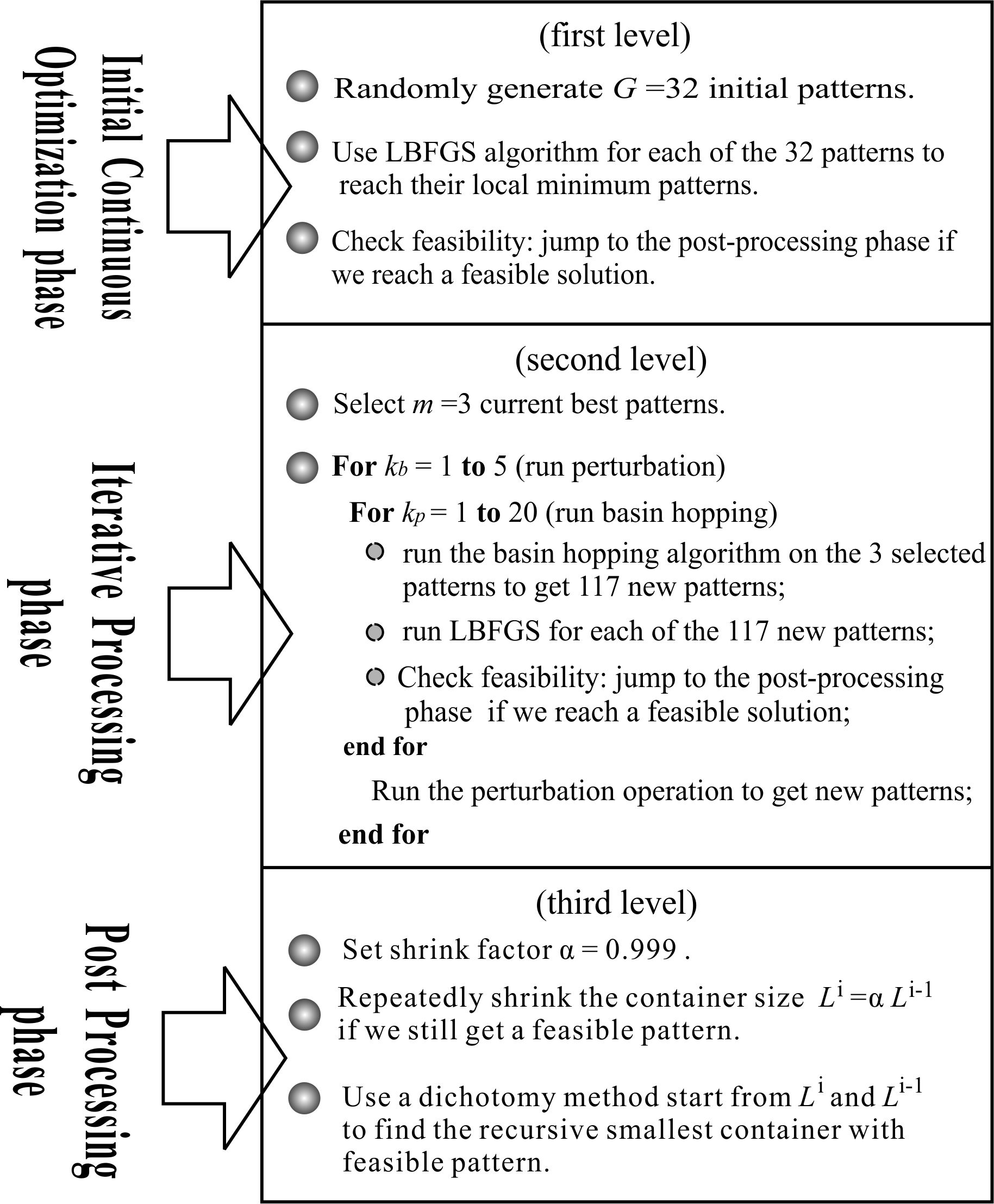}
\caption { The framework of PAC-PNC.}
\label{fig:key8}
\end{figure}

The first phase, the initial continuous optimization phase, corresponds to the first level in Fig. \ref{fig:key8}. Its objective is to generate patterns with local minimum potentials and to select good patterns for the iterative processing phase. Parameter \textit{L} represents the container size and is initialized by the current best record $L_{best-current-result}$ published by ASGO. We randomly generate \textit{G} patterns (\textit{G} = 32) to increase the diversification; then, we use LBFGS on each of the patterns for continuous optimization to reach their local minima. We will switch to the post-processing phase if a feasible solution is found.

The iterative processing phase corresponds to the second level in Fig. \ref{fig:key8}. The main strategy is to call the basin-hopping algorithm to jump out of the local minimum trap. We select \textit{m} patterns (\textit{m} = 3) with the lowest total elastic potential energy from the set of patterns ${G_i | 1 \leq i \leq 32}$ and apply the basin-hopping algorithm to the selected patterns to obtain a total of 39\textit{m} =117 new patterns to increase the diversification. And we apply $k_p$ successive iterations ($k_p$ = 20) of basin hopping. At each iteration, we run the basin-hopping algorithm to generate new patterns that inherit some good properties of the old patterns. We then apply LBFGS to reach their local minima. We will check the feasibility at the end of each iteration and switch to the post-processing phase if there exists a feasible pattern. If we execute the basin-hopping algorithm for $k_p = 20$ times and cannot obtain a feasible solution, we then believe that the current patterns cannot be improved using the basin-hopping strategy alone. For the above parameters, we still refer to the ASGO algorithm and make some tiny changes. We therefore modify the perturbation operator proposed by Fu et al. \cite{3}to perturb each of the \textit{m} = 3 selected patterns to jump out of the local minimum trap. The details are shown in the following:
\begin{enumerate}
\item[$Step 1$:]{ Compute the pain degree $P_i$ for each circle $C_i (1 \leq i \leq n)$.}
\item[$Step 2$:]{ Remove all circles in sets $S_1$ and $S_2$ that contain smaller circles.}
\item[$Step 3$:]{ Sort all circles in $S_3$ and $S_4$ by their pain degrees in non-ascending order to obtain a list {B}.}
\item[$Step 4$:]{ Swap pair-wise circles of B[\textit{i}] and B[\textit{i}+1], where $i = 1, 3, 5, 7, \dots$}
\item[$Step 5$:]{ For all circles in $S_1$ and $S_2$, start from the most painful circle in the old pattern and place each circle in the best-matched action space.}
\item[$Step 6$:]{ Apply the LBFGS algorithm for continuous optimization.}
 \end{enumerate}
 
The perturbation operator strategy will guide the patterns to some promising areas by constructing an updated pattern instead of destroying the pattern. Then, we run the 20 iterations of basin hopping plus continuous optimization again. The outer iteration with the perturbation will run for at most $k_b = 5$ times. If we still cannot find a feasible solution in $k_b$ attempts, we consider that the initial patterns make it difficult to reach a feasible solution and restart the whole process.

The post-processing phase, or the final phase, corresponds to the third level in Fig. \ref{fig:key8}. If we obtain a feasible solution from the above two phases, we then use the post-processing strategy to further reduce the container size while maintaining the feasibility to improve the solution. The details are shown in Algorithm \ref{alg1}.

 \begin{algorithm}[htbp]
 \SetKwInOut{Input}{input}\SetKwInOut{Output}{output}
 \Input{Container size, item sizes $r_1, \dots, r_n$, number of patterns \textit{G}, selection number \textit{m};}
 \Output{Feasible pattern ($X, L$) or unfeasible pattern Null}
 \BlankLine
 \Repeat{reach the limitation of time}{
 \emph {Randomly generate \textit{G} patterns and store in array A}\; 
 \For{ each pattern A[\textit{i}]}
 {
      \If{ $U_e$(A[\textit{i}]) $ < 10^{-20}$}{
         (A[\textit{i}],\textit{L})$\gets$post-processing algorithm (A[\textit{i}],\textit{L})\;
         return (A[\textit{i}],\textit{L})\;}
 } 
 \em{ $k_b\gets0$}\;
 \Repeat {$k_b > 5 $}{
    \For{$k_p=1$ \KwTo $20$}{
       \emph { Select \textit{m} patterns with the lowest $U_e$, for each selected pattern, generate 39 new patterns using the basin-hopping algorithm} \;
        \For{each of the new 39*{\textit{m}} patterns}{
              \emph{$(X',L)\gets$ LBFGS$(X,L)$} \; 
              \If{ $U_e(X')  < 10^{-20}$}{
               $(X',L)\gets $ post-processing algorithm $(X',L)$\;
              return $(X',L)$\;
               }
         }
     }
      \emph{Select \textit{m} patterns with the lowest $U_e$, do perturbation operator, $k_b\gets k_{b+1}$}\;    
 }}
  return Null
  \label{alg1}
 \caption{PAS-PCI algorithm.}
 \end{algorithm}
\subsubsection{Basin-hopping algorithm}
\label{sec3:2:2}
The basin-hopping strategy attempts to relocate some circular items in the unoccupied spaces to generate new and better patterns. Based on the partitioned action space and partitioned circle items, we adopt several strategies to jump out of the local minimum trap for the stuck pattern. Our idea is based on the robust complementary relationship between the small circles and the large unoccupied spaces.

Our main basin-hopping strategy consists of the Neighbour Action Space and the Partitioned Circle Items (PCI). In particular, we select \textit{m} patterns (\textit{m} = 3) with the lowest $U_e$ to generate diverse patterns and partition the \textit{n} circles into four sets $(S_1, S_2, S_3, S_4)$ according to their radii. For each set, we choose a circle item with the highest pain degree $P_i$ and place that item in the largest action space, a matching action space, or a random action space from the two action space lists $(l_1, l_2)$ such that its centre coincides with the centre of the action space. For a narrow action space, we partition the long side to obtain two equally sized action spaces and select two items with the highest pain degree from $S_1$ or one from $S_1$ and the other from $S_2$. We then perform Neighbour Space Occupying in the largest narrow action space, a matching narrow action space or a random narrow action space. This strategy develops a pattern to patterns of a wide extent while simultaneously maintaining the merits of previous patterns. It also helps prevent local cycling because we choose the most painful circles from each group instead of the most painful circles overall, thus increasing the diversity. We also tabu the jammer of each group for one step to prevent further cycling.

Furthermore, to enhance the diversification, we combine several other basin-hopping strategies to generate new patterns, such as randomly swapping pair-wise similar circle items in each set of $(S_1, S_2, S_3, S_4)$. By "similar", we mean that when circles of each set $(S_1, S_2, S_3, S_4)$ are stored according to their radii, two adjacent circles are of similar radii. All of these strategies drive the search process to explore new areas in the solution space. The details of the basin-hopping algorithm are shown in Algorithm \ref{alg2}.
\begin{algorithm}[htbp]
 \SetKwInOut{Input}{input}\SetKwInOut{Output}{output}
 \Input{\textit{m} selected pattern;}
 \Output{For each of the \textit{m} patterns, a list of new patterns is generated as follows:}
 \BlankLine
 {Generate 24 new patterns by choosing one not tabued circle item from each set $(S_1, S_2, S_3, S_4)$ with the highest pain degree and remove it from the square container. For each of the four selected items,
\begin{enumerate}
\item[(1)]{ Place it into the largest action space in list $l_1$, and obtain four new patterns. }
\item[(2)]{Place it into the largest action space in list $l_2$, and obtain four new patterns.} 
\item[(3)]{ Place it into the best matching action space in list $l_1$ or $l_2$, and obtain eight new patterns.}
\item[(4)]{ Place it into a randomly selected action space in list $l_1$ or $l_2$, and obtain eight new patterns.}
\end{enumerate}}
{ Generate three new patterns by choosing two not tabued circle items from set $S_1$ with the highest pain degree, and then
\begin{enumerate}
\item[(1)]{ Perform Neighbour Space Occupying (NSO) on the largest Narrow Action Space (NAAS) to obtain a new pattern.}
\item[(2)]{ Select the best matching NAAS for the most painful circle item in $S_1$ and perform NSO to obtain a new pattern.}
\item[(3)]{ Perform NSO on a random NAAS to obtain a new pattern.}
 \end{enumerate} }
{ Generate three new patterns by choosing two not tabued circle items with the greatest pain degree, one from $S_1$ and another from $S_2$, and then
\begin{enumerate}
\item[(1)]{ Perform NSO on the largest NAAS to obtain a new pattern.}
\item[(2)]{ Choose the best matching NAAS for the selected item from $S_1$ and perform NSO.} 
\item[(3)]{ Choose the best matching NAAS for the selected item from $S_2$ and perform NSO.}
\end{enumerate}     }
{ Generate five new patterns as follows:
\begin{enumerate}
\item[(1)]{ Randomly swap pair-wise similar circles of $S_1$ with $S_2$.}
\item[(2)]{ Randomly swap pair-wise similar circles of $S_2$ with $S_3$.}
\item[(3)]{ Randomly swap pair-wise similar circles of $S_3$ with $S_4$.}
\item[(4)]{ Randomly swap pair-wise similar circles of $S_1$ with $S_3$.}
\item[(5)]{ Randomly swap pair-wise similar circles of $S_2$ with $S_4$.}
\item[]{For random swap of ${C_i, C_i+1}$ of a set and ${C_j, C_{j+1}}$ of another set, we mean swap $C_i$ and $C_j$ and swap $C_{i+1}$ and $C_{j+1}$.}
\end{enumerate}}
For each set of $(S_1, S_2, S_3, S_4)$, select a circle item $C_i$ with the greatest pain degree $P_i$ and swap $C_i$ with $C_{i+1}$. Thus, obtain four new patterns.
\BlankLine
For each set of $(S_1, S_2, S_3, S_4)$, randomly swap two circle items. Thus, obtain four new patterns.
\caption{Basin-hopping algorithm.}
\label{alg2}
\end{algorithm}
\subsubsection{Post-processing algorithm}
\label{sec3:2:3}
Finally, the PAS-PCI algorithm calls for post-processing of a feasible pattern to find a better feasible solution with a smaller container size. The post-processing algorithm repeatedly shrinks the container size by a constant factor of $\propto = 0.999$ until a feasible pattern can no longer be obtained. Then, a dichotomy method is applied to find the best feasible solution. Algorithm \ref{alg3} shows the details of the post-processing algorithm.

\begin{algorithm}[htbp]
    \SetKwInOut{Input}{input}\SetKwInOut{Output}{output}
   \Input{A feasible solution ($X, L$), shrinking factor $\propto$;}
   \Output{A feasible solution, in which $L' \leq L$;}
\Repeat{$ U_e(X') > 10^{-20}$}{
	 $ L = \propto L$\;
	 $ X'\gets$ LBFGS ($X,L$)\;
}
    $L_{upper}\gets L/\propto,$ $L_{lower}\gets L$ \;
\Repeat{$L_{upper}- L_{lower} <〖10^ {-7}$}{
        $ L'\gets(L_upper + L_lower)/2$\;
	    $ X'\gets$ LBFGS $(X,L')$\;
        \eIf{$U_e(X') < 10^{-20}$}{
     	     $ L_{upper}\gets L'$\;
     	}{     
             $ L_{ower}\gets L'$\;}
} 
   \caption{Post-processing algorithm.}
   \label{alg3}
\end{algorithm}

\section{Computational results }
\label{sec4}
To assess the efficiency of the proposed approach, we implemented PAS-PCI in the Visual C++ programming language. We ran PAS-PCI on a personal computer with 3.0 GHz and 4.0 GB memory.

We tested PAS-PCI on two groups of benchmark instances: $r_i=i$ and $r_i=
\sqrt{i}$ ($r_i$ is the radius of circle i, $1 \leq i \leq n$). The instances for $r_i=i$ are large-variation instances, and the second group $r_i=\sqrt{i}$)  contains small-variation instances. On the Packomania website, the range of \textit{n} is from 1 to 72. We executed each instance 10 times to reduce the effects of randomness and find the best results compared with the ASGO algorithm. For each instance, the search process will terminate if PAS-PCI yields a feasible solution no worse than the current best results or if it reaches the time limit. We compared PAS-PCI with both our previous Action Space based Global Optimization (ASGO) algorithm \cite{5} and the best results downloaded from the Packomania website, which were obtained by different authors \cite{27}. The new records obtained in these experiments are published on www.packomania.com by Eckard Specht \cite{27}. He also has further optimized our originally obtained results using some exchange heuristics and relocations \cite{26}. In general, PAS-PCI generates more-satisfying solutions.

\subsection{Parameter setup}
\label{sec4:1}
\begin{figure}[htbp]
\centering
\includegraphics[scale=0.5]{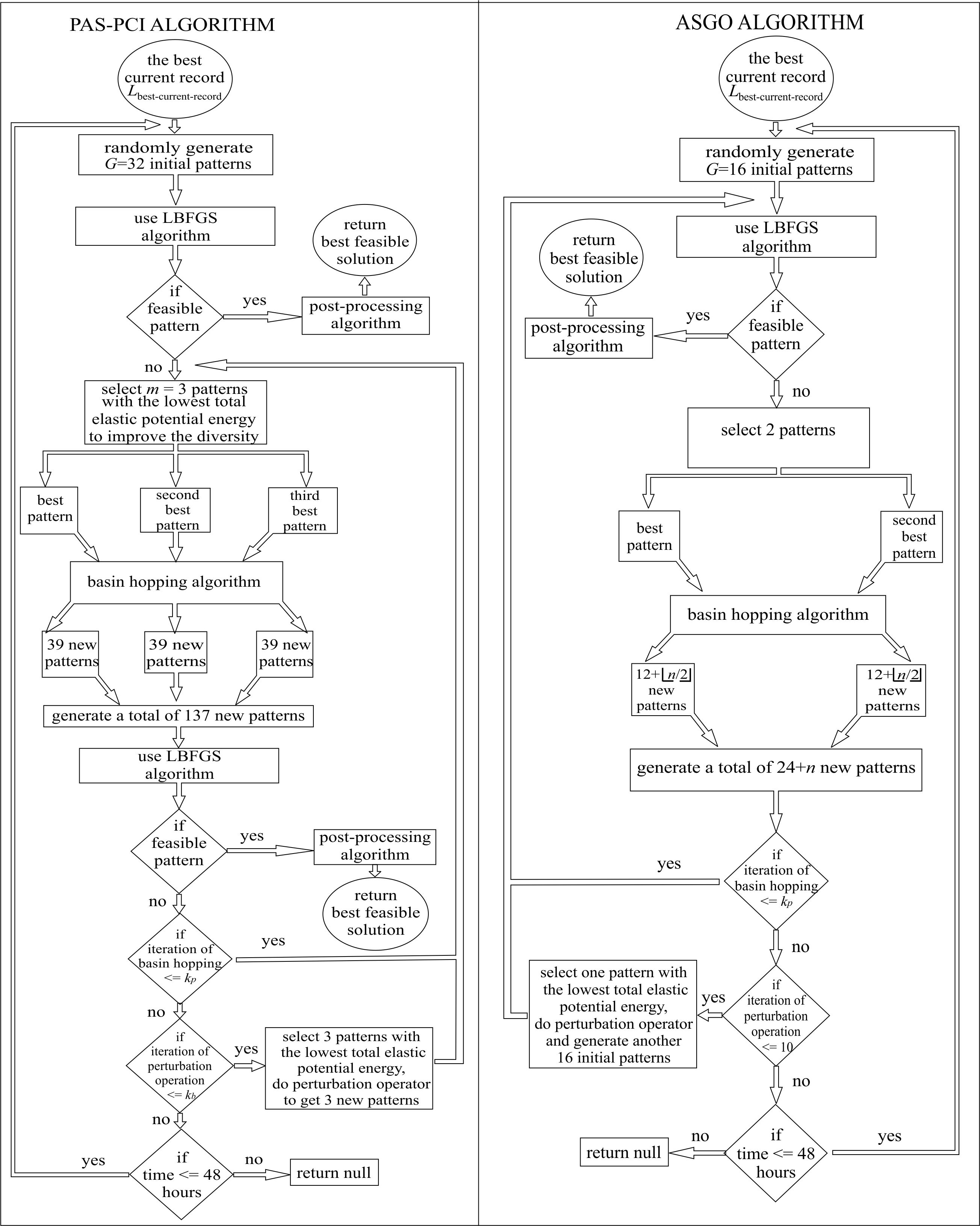}
\caption { The parameter setup in the flow of PAS-PCI and ASGO. }
\label{fig:key9}
\end{figure}

All of the parameters used in the PAS-PCI algorithm are shown in Table \ref{tab:1}, selected based on a small number of trials or referred to parameters of the previous algorithm ASGO. In Table \ref{tab:2}, we give two examples to show the impact of parameter \textit{m}, the number of selected patterns with the lowest elastic potential energy for basin hopping. Here we only show the evaluations on sampled instances of \textit{n} = 27, and \textit{n} = 64 for the $r = i$ benchmark set. We run PAS-PCI on each instance for 5 times (\textit{m} = 1, 2, 3, 4, 5) and record the results. For each selected value of \textit{m}, we show the number of hit times for 5 trials, and the average running times for successful trails. We also list the best, worst and average solutions. We could see that \textit{m} = 3 yield the best performance. For other key parameters, we also fix other parameters and do small number of trials to select the appropriate value.
Table \ref{tab:1} provides a brief summary of the parameters used in the algorithm. In the basin hopping process, the circles are sorted based on their sizes, and they are evenly partitioned into four groups $S_1$ to $S_4$ based on their radii. $S_1$ and $S_2$ contain smaller circles and they have higher priority for the movement, and we tabu the jammer of each group for one step to prevent the cycling. The container size \textit{L} is initialized by the current best record $L_{best-current-result}$, and the quality of a feasible solution is measured by the final \textit{L} for each instance. $l_1$ and $l_2$ are two lists, each list contains the top 10 action spaces depending on the evaluation metric.
Fig. \ref{fig:key9} illustrates the comparison of the parameters for PAS-PCI and ASGO \cite{5} in the framework. In previous work, He et al. \cite{5} generated 16 initial patterns and choose 2 patterns at the end of each LBFGS iteration. To increase the diversity, we double the initial number of random patterns to \textit{G} = 32, and select \textit{m} = 3 patterns with the lowest elastic potential energy at the end of each LBFGS iteration. Then in the following basin hopping process, we will reach a total of 39\textit{m} = 117 new patterns. For every $k_p$ = 20 iterations of basin hopping, He et al. \cite{5} use a perturbation operator to perturb the patterns to some extent, and their outer iterations for a perturbation run for $k_b$ = 10 loops. In the following experiments that $n\in[14,72]$, as we generate considerably more new patterns in the inner iteration, we set $k_b$ = 5 to save the total running time. Finally we use the same time limit of 48 hours to compare with ASGO \cite{5}.

\begin{table}[htbp]
\caption{Parameter regulation.} 
\label{tab:1}
\centering 
\begin{tabular}{l l l} 
\hline 
Parameter  &	Description	& Value\\ [0.5ex] 
\hline 
\textit{n}  &	Number of circle items  &	14-72\\
$C_i$       &	The $i^th$ circle item	\\
\textit{L}	& Size of the square container & Initialized by the current-best-result\\
\textit{X}	& Coordinates for all centres of the circles & $(x_1, y_1, \dots, x_n, y_n)$\\
$S_1$   & Set of circles            &	Circles index in $[1, int (n/4)])] $ \\
$S_2$       & Set of circles &	Circles index in [int (\textit{n}/4) +1, int (\textit{n}/2)]\\
$S_3$ &	Set of circles & Circles index in [int (\textit{n}/2)+1, int(3\textit{n})/4]\\
$S_4$ & Set of circles & Circles index in [int (3\textit{n}/4) +1, \textit{n}]\\
\textit{N}	 & Tabu tenure & 	1 \\
$l_1, l_2$ &	Lists of largest action spaces &	$|l_1 |=|l_2 |=10$ \\
\textit{G} &	Number of random patterns &	32\\
\textit{m} &	Number of initial patterns for basin hopping &	3 \\
$k_p$ &	Number of successive iteration &	20 \\
$k_b$ &	Perturbations &	5 \\
B[\textit{i}] &	The circle in the $i^th$ position of list B	\\
\textit{T} &	Allowed running time &	48 h \\
\hline 
\end{tabular}
\end{table}

\begin{table}[htbp]
\caption{Evaluations on parameter m for PAS-PCI (sampled on n = 27, 64 and m = 1, 2, 3, 4, 5).}
\label{tab:2}
\centering
\begin{tabular}{c }
\textit{n}=27
\end{tabular}
\begin{tabular}{l c c c c c }
\hline\noalign{\smallskip}
\textit{n} & Hits &	Average running & Best solution & Worst solution &	Average solution \\ 
& & time \\
\noalign{\smallskip}\hline\noalign{\smallskip}
1 &	2/5	& 16,751 &	159.39844572 &	159.40009811 &	159.399271915 \\
2 &	3/5 & 13,634 &	159.25909220 &	159.39642779 &	159.316007847 \\
3 &	5/5	& 3,215	 &  159.10320551 &	159.26084334 &	159.206795856 \\
4 &	5/5	& 6,908	 &  159.18019927 &	159.30947629 &	159.246277900 \\
5 &	5/5	& 8,216	 &  159.16913324 &	159.27609935 &	159.238197052 \\
\noalign{\smallskip}\hline
\end{tabular}
\begin{tabular}{c }
\textit{n}=64
\end{tabular}
\begin{tabular}{l c c c c c }
\hline\noalign{\smallskip}
\textit{n} & Hits &	Average running & Best solution & Worst solution &	Average solution \\ 
& & time \\
\noalign{\smallskip}\hline\noalign{\smallskip}
1 &	1/5 & 90,454 &	568.46130293 &  568.46130293 &	568.46130293 \\
2 &	2/5 & 76,479 &	568.40597817 &	568.45627099 &	568.43112458 \\
3 &	3/5 & 61,537 &	568.37716941 &	568.41008273 &	568.38910549 \\
4 &	3/5 & 69,862 &	568.37716941 &	568.43158391 &	568.40303589 \\
5 &	3/5 & 73,058 &	568.38011068 &	568.42011171 &	568.39874129 \\
\noalign{\smallskip}\hline
\end{tabular}
\end{table}

\subsection{Comparison on $r_i=i$}
\label{sec4:2}
We compared our results on the benchmark instances of $r_i=i$ with two works: ASGO \cite{5} and the results on the Packomania website \cite{27}. The number of circle items per instance that ASGO executed ranges from 15 to 45, in addition to instances 50, 55 and 60. We compared all 34 benchmark instances that ASGO executed with the proposed PAS-PCI algorithm. PAS-PCI returned feasible patterns with smaller container sizes in 33 instances and matched one other instance; PAS-PCI also ran faster in most instances.

We then compared the record results for \textit{i} = 14 to 72 on the Packomania website obtained by several researchers. PAS-PCI returned smaller container sizes in 50 instances and matched the other 9 instances. Note that we updated 19 records for the set of $r_i=i$ (47-48, 51-54, 57, 61-72) that had been kept unchanged since May 21, 2013.
\subsection{Comparison on $r_i=\sqrt{i}$}
 \label{sec4:3}
We also compared our results to the instances for $r_i=\sqrt{i}$. For ASGO, the range of ASGO instances started from 15 to 45, in addition to the instances 50, 55 and 60. We compared all of the 34 benchmark instances that ASGO executed with the proposed PAS-PCI algorithm. PAS-PCI returned feasible patterns with smaller container sizes in 31 instances and matched the other three; PAS-PCI also ran faster in most instances. Comparison of 35 instances (from 14 to 45, in addition to instances 50, 55 and 60 for larger instances) from the Packomania website with the PAS-PCI results shows that PAS-PCI returned feasible patterns with smaller container sizes in 32 instances and matches for the other 3 instances.

\begin{figure}[htbp]
\centering
\includegraphics[scale=0.32]{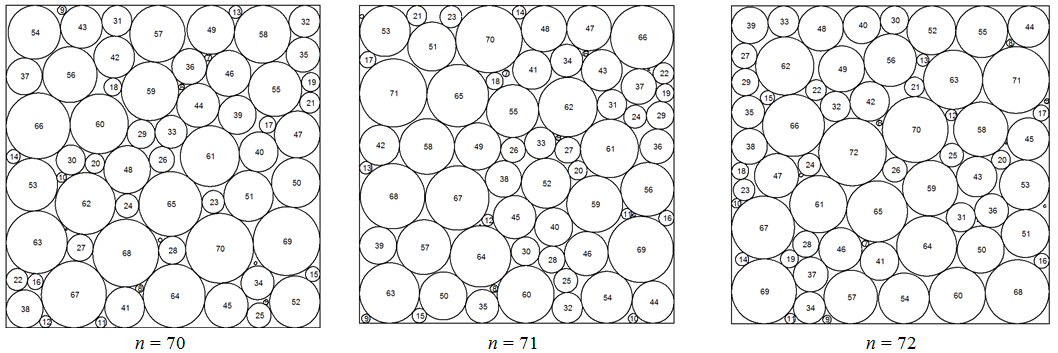}
\caption { Packing layouts for $r_i=i$ }
\label{fig:key10}
\end{figure}

\begin{figure}[htbp]
\centering
\includegraphics[scale=0.31]{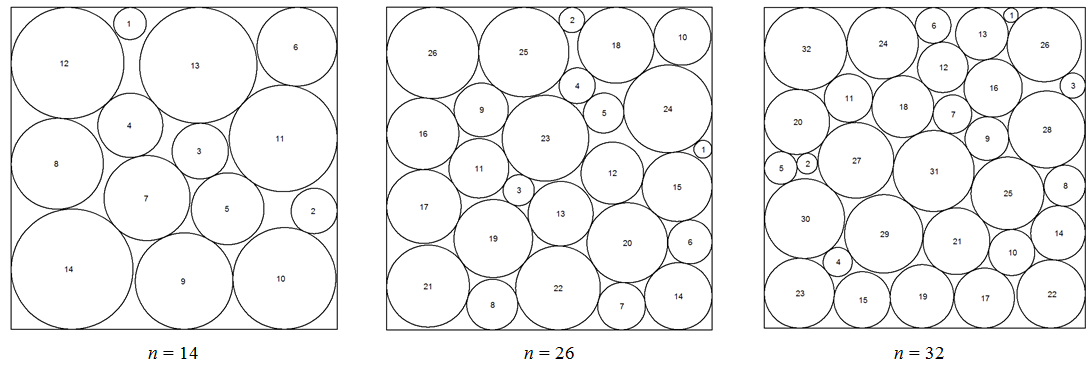}
\caption { Some better packing layouts for $r_i=\sqrt{i} $ }
\label{fig:key11}
\end{figure}

Tables \ref{tab:3} and \ref{tab:4} show the computational results for PAS-PCI, ASGO and the Packomania website for $r_i=i$ and $r_i=\sqrt{i}$, respectively. The first column includes the lists of the number of circles, the second column includes the lists of the best-known results obtained by the ASGO algorithm, and the third column includes the lists of the record results we downloaded from the Packomania website. Column 4 includes the best solutions of PAS-PCI, columns 5-8 show the best running time and average running time for ASGO and PAS-PCI, and columns 9 and 10 include the number of hit times for 10 tests on ASGO and PAS-PCI, respectively. With the new results of the instances, several of the new forms can be seen in Fig. \ref{fig:key10} and \ref{fig:key11}.

\begin{table}[htbp] 
\caption{Computational results for PAS-PCI, ASGO and the Packomania website when $r_i=i.$}
\label{tab:3}
\centering
\begin{tabular}{l l l l l l l l c c}
\hline\noalign{\smallskip}
 &	Best-known  & Best-known  & Best-solution  & Best-time & Best-time & Average time & Average time & Hits  &	Hits \\
 \textit{n}       & ASGO        & Packomania  & PAS-PCI        & ASGO (s)      & PAS-PCI(s) & ASGO (s) &  PAS-PCI (s) & ASGO  & PAS-PCI \\
\noalign{\smallskip}\hline\noalign{\smallskip}
14 & -----------  & 61.84992131  & 61.84992131  & -----   & 39      & -----   &	73      & ----- & 10/10\\
15 & 68.52756391  & 68.52756391  & 68.52756391  & 101     & 29      & 106     &	86      & 10/10 & 10/10 \\
16 & 75.04786163  & 75.00934256  & 75.00934256  & 37	  & 35      & 582     & 275     & 10/10 & 10/10 \\
17 & 81.50268767  & 81.50268767  & 81.50229942  & 111     &	49      & 956     & 543     & 10/10 & 10/10 \\
18 & 88.70023386  & 88.40408812  & 88.40408757  & 290     &	93      & 1,844   & 712     & 10/10 & 10/10 \\
19 & 95.78524607  & 95.96921334  & 95.77583449  & 36	  & 121     & 3,133   & 1,563   & 10/10 & 10/10 \\
20 & 104.92138708 & 103.11765325 & 103.11765325 & 3       & 105     & 96	  & 248     & 10/10 & 10/10 \\
21 & 110.56074443 &	110.56029719 & 110.56029719 & 1,039	  & 194     & 3,885   & 2,477   & 10/10 & 10/10 \\
22 & 118.61665487 &	118.65797248 & 118.60607058 & 491     & 261     & 2,707   & 1,591   & 10/10 & 10/10 \\
23 & 126.34478973 &	126.49109690 & 126.19548212 & 73      & 47      & 1,793   & 2,963   & 10/10 & 10/10 \\
24 & 134.37268240 &	134.36183861 & 134.28692283 & 610     & 129     & 5,531   & 2,148   & 10/10 & 10/10 \\
25 & 142.62630934 &	142.67641929 & 142.53983356 & 372     & 186     & 2,874   & 907     & 10/10 & 10/10 \\
26 & 151.15018638 &	151.05511016 & 150.99258664 & 759     & 274     & 2,546   & 1,862   & 10/10 & 10/10 \\
27 & 159.40352999 &	159.43924161 & 159.10320551 & 1,800	  & 465	    & 14,979  & 3,314   & 10/10 & 10/10 \\
28 & 168.35062099 &	168.38155796 & 168.09364893 & 599	  & 301	    & 3,627	  & 2,394	& 10/10 & 10/10 \\
29 & 177.09268466 &	177.15283242 & 176.99373443 & 3,358	  & 826	    & 9,883	  & 5,468	& 10/10 & 10/10 \\
30 & 186.12550501 &	186.10273226 & 185.97476319 & 457	  & 719	    & 20,236  & 10,531  & 10/10 & 10/10 \\
31 & 195.30022174 &	195.36206001 & 195.21263867 & 1,272	  & 1,028   & 9,335	  & 7,818	& 10/10 & 10/10 \\
32 & 204.45658766 &	204.53949877 & 204.16829349 & 3,337	  & 485	    & 21,536  & 8,604	& 10/10 & 10/10 \\
33 & 214.22777540 &	214.38380598 & 214.19767547 & 581	  & 1,553   & 7,960	  & 11,317  & 10/10 & 10/10 \\
34 & 224.10619117 &	223.84825302 & 223.77519778 & 429	  & 291	    & 2,696	  & 1,928	& 10/10 & 10/10 \\
35 & 233.51807964 &	233.37623193 & 233.10099573 & 2,737	  & 946     & 13,183  & 10,396  & 10/10 & 10/10 \\
36 & 242.99143037 &	242.96247182 & 242.71634934 & 3,701	  & 2,764   & 17,636  & 8,414	& 10/10 & 10/10 \\
37 & 253.62316424 &	253.61341154 & 253.19632297 & 1,215	  & 1,129   & 3,951	  & 12,937  & 10/10 & 10/10 \\
38 & 263.53652205 &	263.58677860 & 262.89961522 & 931	  & 576     & 12,798  & 7,328	& 10/10 & 8/10 \\
39 & 273.75141478 &	273.71426330 & 273.31495736 & 502	  & 1,832   & 9,137	  & 6,485	& 10/10 & 9/10 \\
40 & 284.30264444 &	284.21598927 & 283.79885749 & 1,877	  & 941	    & 9,339	  & 8,952   & 10/10 & 8/10 \\
41 & 293.89806626 &	293.78879283 & 293.78100971 & 1,555	  & 3,092   & 5,850	  & 15,329  & 4/10  & 7/10 \\
42 & 304.80698321 &	304.78182216 & 304.72298974 & 3,433	  & 2,463   & 10,126  & 8,594	& 6/10  & 5/10 \\
43 & 315.61313564 &	315.59938719 & 315.51991431 & 7,843	  & 6,097   & 33,285  & 21,319  & 5/10  & 8/10 \\
44 & 327.24688870 &	326.95644304 & 326.51876564 & 926	  & 10,043  & 6,842	  & 25,377  & 10/10 & 9/10 \\
45 & 338.08419513 &	337.93815812 & 337.63639881 & 2,859	  & 1,993   & 9,743	  & 9,301   & 10/10 & 6/10 \\
46 & ------------ &	349.05075115 & 349.01221399 & -----	  & 7,309   & -----	  & 21,865  & ----- & 8/10 \\
47 & ------------ &	360.33649558 & 360.28848158 & -----	  & 2,942   & -----	  & 23,459  & ----- & 8/10 \\
48 & ------------ &	371.24394236 & 371.14228033 & -----	  & 11,916  & -----	  & 19,141  & ----- & 6/10 \\
49 & ------------ &	382.35123818 & 382.35123818 & -----	  & 43,604  & -----	  & 96,244  & ----- & 7/10 \\
50 & 394.15415832 &	394.11379133 & 393.99311849 & 20,397  & 5,637   & 121,061 & 31,493  & 4/10  & 5/10 \\
51 & ------------ &	405.98821870 & 405.95674675 & -----	  & 14,394  & -----	  & 48,151  & ----- & 4/10 \\
52 & ------------ &	418.76983792 & 418.76983293 & -----	  & 18,557  & -----	  & 64,878  & ----- & 7/10 \\
53 & ------------ &	429.56613726 & 429.48485256 & -----	  & 1,693   & -----	  & 6,565   & ----- & 10/10 \\
54 & ------------ &	441.70943495 & 441.63484419 & -----	  & 18,031  & -----	  & 65,028  & ----- & 9/10 \\
55 & 453.27302397 &	453.23743150 & 452.98196941 & 134,338 &	20,197  & 142,763 & 73,761  & 2/10  & 6/10 \\
56 & ------------ &	466.65243154 & 466.65243154 & -----	  & 26,946  & -----	  & 61,915  & ----- & 2/10 \\
57 & ------------ &	479.14731853 & 479.14003521 & -----	  & 22,364  & -----	  & 34,921  & ----- & 5/10 \\
58 & ------------ &	490.74893276 & 490.74893276 & -----	  & 32,108  & -----	  & 59,991  & ----- & 4/10 \\
59 & ------------ &	502.95832470 & 502.95832470 & -----	  & 29,398  & -----	  & 68,242  & ----- & 7/10 \\
60 & 516.47881793 &	516.61305890 & 516.39141391 & 21,227  & 19,628  & 88,476  & 42,238  & 10/10 & 5/10 \\
61 & ------------ &	529.99890873 & 529.91176839 & -----	  & 15,608  & -----	  & 42,739  & ----- & 10/10 \\
62 & ------------ &	543.31155400 & 542.94950697 & -----	  & 9,993   & -----	  & 63,716  & ----- & 10/10 \\
63 & ------------ &	555.09335338 & 555.09322952 & -----	  & 3,492   & -----	  & 31,836  & ----- & 3/10 \\
64 & ------------ &	568.40130293 & 568.37716941 & -----	  & 18,011  & -----	  & 56,352  & ----- & 6/10 \\
65 & ------------ &	581.17833643 & 580.71949298 & -----	  & 4269    & -----	  & 32,981  & ----- & 8/10 \\
66 & ------------ &	595.94350942 & 595.93622413 & -----	  & 3719    & -----	  & 17,698  & ----- & 10/10 \\
67 & ------------ &	609.15441946 & 609.15441108 & -----	  & 13,464  & -----	  & 50,492  & ----- & 4/10 \\
68 & ------------ &	622.42276086 & 622.42125971 & -----	  & 12,293  & -----	  & 37,963  & ----- & 8/10 \\
69 & ------------ &	634.31384012 & 634.30215215 & -----	  & 27,068  & -----	  & 108,762 & ----- & 8/10 \\
70 & ------------ &	648.03770190 & 647.87846558 & -----	  & 39,457  & -----	  & 126,511 & ----- & 8/10 \\
71 & ------------ &	662.83805941 & 662.83401285 & -----	  & 52,603  & -----	  & 148,681 & ----- & 2/10 \\
72 & ------------ &	677.34672378 & 677.33531058 & -----	  & 19,837  & -----   & 88,927  & ----- & 5/10 \\
\noalign{\smallskip}\hline
\end{tabular}
\end{table}

\begin{table}[htbp]
\caption{Computational results for PAS-PCI, ASGO and the Packomania website when $r_i=\sqrt{i}.$}
\label{tab:4}  
\centering     
\begin{tabular}{l l l l l l l l c c}
\hline\noalign{\smallskip}
 &	Best-known  & Best-known  & Best-solution  & Best-time & Best-time & Average time & Average time & Hits  &	Hits \\
 \textit{n}       & ASGO        & Packomania  & PAS-PCI        & ASGO (s)      & PAS-PCI(s) & ASGO (s) &  PAS-PCI (s) & ASGO  & PAS-PCI \\
\noalign{\smallskip}\hline\noalign{\smallskip}

14 &	----------- &	20.03404652  &	20.03384653 &	-----  &	34     & -----   &	49     &	----  &	10/10 \\
15 &	21.29813169 &	21.29813169  &	21.29813169 &	101    &	37     & 275     &	139    &	10/10 &	10/10 \\
16 &	22.56309981 &	22.56309981  &	22.56309981 &	48     &	46     & 2,657   &	186    &	10/10 &	10/10 \\
17 &	23.87238984 &	23.87238984  &	23.87238984 &	1,201  &	193	   & 4,392   &	549    &	10/10 &	10/10 \\
18 &	25.29143658 &	25.30990723  &	25.27915927 &	215    &	135	   & 1,092   &	809    &	10/10 &	10/10 \\
19 &	26.64511657 &	26.66753375  &	26.62601755 &	424    &	183	   & 1,734   &	725    &	10/10 &	10/10 \\
20 &	27.99580859 &	27.99556516  &	27.97923319 &	126    &	154	   & 2,044   &	1,312  &	10/10 &	10/10 \\
21 &	29.33636969 &	29.37295726  &	29.30269792 &	219    &	186	   & 1,528   & 	1,286  &	10/10 &	10/10 \\
22 &	30.73679088 & 	30.74179521  &	30.09246437 &	382    &	182	   & 1,244   &	604    &	10/10 &	10/10 \\
23 &	32.02058355 &	32.03166165  &	32.00872179 &	2,200  &	1,019  & 6,229   &	2,964  &	10/10 &	10/10 \\
24 &	33.38242107 &	33.41437096  &	33.36991643 &	303    &	434	   & 1,909   &	1,653  &	10/10 &	10/10 \\
25 &	34.69652756 &	34.70564553  &	34.69392717 &	1,044  &	2,307  & 6,835   &	5,816  &	10/10 &	10/10 \\
26 &	36.05840859 &	36.07166003  &	36.02915819 &	981    &	794	   & 3,161   &	1,768  &	10/10 &	10/10 \\
27 &	37.32721394 &	37.35117334  &	37.31109810 &	6,268  &	962	   & 28,651  &	4,851  &	10/10 &	10/10 \\
28 &	38.68609137 &	38.69314892  &	38.62981649 &	16,538 &	3,843  & 28,586  &	4,327  &	10/10 &	10/10 \\
29 &	40.05783844 &	40.07711813  &	40.00191663 &	1,390  &	2,429  & 15,224  &	9,743  &	10/10 & 10/10 \\
30 &	41.42606085 &	41.45220322  &	41.38741929 &	2,917  &	1,846  & 7,565   &	3,012  &	10/10 &	10/10 \\
31 &	42.75687834 &	42.793380560 &	42.70018672 &	1,781  &	1,983  & 7,348   &	6,907  &	10/10 & 10/10 \\
32 &	44.04766016 &	44.063333662 &	44.03950493 &	2,728  &	4,212  & 52,999  &	15,471 &	10/10 &	10/10 \\
33 &	45.38587795 &	45.394745307 &	45.37281453 &	19,864 &	9,305  & 60,859  &	23,549 &	10/10 &	10/10 \\
34 &	46.75308313 &	46.752000809 &	46.71956130 &	6,204  &	3,189  & 33,474  &	18,681 &	10/10 &	10/10 \\
35 &	48.06709480 &	48.073923426 &	48.06074066 &	16,541 &	6,948  & 52,676  &	14,934 &	4/10  &	10/10 \\
36 &	49.40082616 &	49.400079475 &	49.37011019 &	9,568  &	5,513  & 17,251  &	9,626  &	2/10  &	10/10 \\
37 &	50.81979234 &	50.827145108 &	50,79142199 &	2,784  & 	4,367  & 7,033   &	19,238 &	10/10 &	10/10 \\
38 &	52.14070553 &	52.150810546 &	52.06979691 &	485    &	3,753  & 15,734  &	8,625  &	10/10 &	10/10 \\
39 &	53.44302697 &	53.473053122 &	53.37364937 &	3,995  &	2,364  & 21,022  &	17,682 &	10/10 &	10/10 \\
40 &	54.80450469 &	54.809098043 &	54.78167521 &	1,784  &	903	   & 10,028  &	7,108  &	10/10 &	9/10 \\
41 &	56.19655477 &	56.215141936 &	56.15964905 &	3,649  &	2,705  & 9,670   &	11,754 &	10/10 &	9/10 \\
42 &	57.47214913 &	57.484131297 &	57.40197437 &	3,406  &	1,983  & 14,909  &	13,421 &	10/10 &	9/10 \\
43 &	58.80263818 &	58.802206944 &	58.78746261 &	19,060 &	7,936  & 50,342  &	15,019 &	3/10  &	7/10 \\
44 &	60.20692438 &	60.210820103 &	60.15943733 &	4,821  &	9,208  & 21,833  &	18,229 &	10/10 &	7/10 \\
45 &	61.44936316 &	61.455016469 &	61.40695896 &	32,287 &	11,385 & 66,868  &	22,388 &	2/10  &	6/10 \\
50 &	68.20736878 &	68.231344973 &	68.16295751 &	3,799  &	3,748  & 20,153  &	17,997 &	8/10  &	6/10 \\
55 &	74.96860671 &	74.977762353 &	74.92900891 &	4,271  &	8,394  & 13,307  &	23,929 &	8/10  &	5/10 \\
60 &	81.54674520 &	81.546626989 &	81.51614965 &	85,661 &	34,091 & 106,258 &	52,672 &	2/10  &	4/10 \\

\noalign{\smallskip}\hline
\end{tabular}
\end{table}

\section{Discussion and conclusions}
\label{sec5}
This paper proposes a Partitioning Narrow Action Spaces and Circle Items (PAS-PCI) algorithm for packing unequal circles in a two-dimensional square container (PUC-SC). The effective LBFGS algorithm is applied to access a local minimum depending on the potential energy. To offer the best results of the search process and to improve on previous basin-hopping strategies, we present a new basin-hopping strategy consisting of two main ideas: the partition of the circular items based on their sizes and the partition of the narrow action spaces. We partition the circular items into four sets, $S_1$ to $S_4$, according to their radii from small to large and propose basin-hopping strategies on different sets to prevent local cycling and increase the possibility of finding appropriate unoccupied spaces for the circles with low computational complexity.

We approximate each circle item as a square item and measure the size of an action space in two ways, the length of the short side and the perimeter. We then sort all action spaces in non-ascending order based on the two metrics to order the two lists, which can find more suitable action spaces when moving squeezed circles, particularly small circles. Then, we partition all of the top narrow action spaces in the two lists by dividing each Narrow Action Space at the centre of the long side to produce two equal-size action spaces that we call Neighbour Action Spaces. We select two items with the greatest pain degree from $S_1$, or one from $S_1$ and the other from $S_2$, and perform Neighbour Space Occupying.

Furthermore, to enhance diversification in the process of research, we take the best features of the previous basin-hopping strategies and combine them with a perturbation operator by randomly swapping two large or small circles. Experimental results from two previous works, each of which consist of two sets of instances, show the effectiveness of the proposed method.

In future work, we will apply the idea of the Action Space (AS) and the Narrow Action Space (NAAS) to other circle packing problems, such as packing equal or unequal circles into a circular container. One could also expand our approach to address three-dimensional circle-packing problems.

\begin{acknowledgements}
We are grateful to the anonymous referees for valuable suggestions and comments that have helped us improve the paper. This work was supported by the National Natural Science Foundation of China (Grant no. 61472147, 61370183 and 61602196). We sincerely thank Eckard Specht for processing our results and publishing them on the Packomania website.
\end{acknowledgements}

\end{document}